\documentclass[twocolumn,showpacs,preprintnumbers,amsmath,amssymb]{revtex4}  
\usepackage{graphicx}% Include figure files
\usepackage{dcolumn}% Align table columns on decimal point
\usepackage{bm}% bold math
\usepackage{multirow}
\usepackage{color}
\usepackage{amsmath}
\begin{document}
\title{Exciton dynamics in WSe$_2$ bilayers}

\author{G. Wang}
\author{X. Marie}
\author{L. Bouet}
\author{M. Vidal}
\author{A. Balocchi}
\author{T. Amand}
\author{D. Lagarde}
\author{B. Urbaszek}
%\email{urbaszek@insa-toulouse.fr}
\affiliation{%
Universit\'e de Toulouse, INSA-CNRS-UPS, LPCNO, 135 Av. de Rangueil, 31077 Toulouse, France}
%\listfiles
%\nofiles
%\keywords{WSe2, valley dynamics, time resolved photoluminescence, transition metal dichalcogenides, two dimensional materials}
%\date{\today}

\begin{abstract}
We investigate exciton dynamics in 2$H$-WSe$_2$ bilayers in time-resolved photoluminescence (PL) spectroscopy. 
Fast PL emission times are recorded for both the direct exciton with $\tau_D\lesssim3$~ps and the indirect optical transition with $\tau_I\approx25$~ps. For temperatures between 4 to 150~K  $\tau_I$ remains constant. Following polarized laser excitation, we observe for the direct exciton transition at the \textit{K} point of the Brillouin zone efficient optical orientation and alignment during the short emission time $\tau_D$. The evolution of the direct exciton polarization and intensity as a function of excitation laser energy is monitored in PL excitation (PLE) experiments.
\end{abstract}

%\pacs{78.60.Lc,78.66.Li}

                           %Use showkeys class option if keyword
                             %display desired
\maketitle
Thin layers of transition metal dichalcogenides (TMDCs), such as MoS$_2$, MoSe$_2$, WS$_2$ and WSe$_2$ have emerged as very promising materials for optical, electronic and quantum manipulation applications \cite{Geim:2013a,Xu:2014a}. 
The opto-electronic and spin properties in TMDCs can be controlled at an atomic layer level:  In TMDC \textit{monolayers} (MLs), the lowest energy inter-band transition at typically 1.8~eV is direct in k-space \cite{Mak:2010a,Splendiani:2010a} with strong optical absorption ( $\approx10\%$). In MLs crystal inversion symmetry breaking together with the strong spin-orbit (SO) interaction leads to a coupling of carrier spin and k-space valley physics, i.e., the circular polarization ($\sigma^+$ or $\sigma^-$) of the absorbed or emitted photon can be directly associated with selective carrier excitation in one of the two non-equivalent $K$ valleys ($K^+$ or $K^-$, respectively) \cite{Xiao:2012a, Cao:2012a,Mak:2012a,Sallen:2012a,Kioseoglou:2012a,Jones:2013a,Mak:2014a}. Moreover, the strong Coulomb interaction between electrons and holes results in large exciton binding energies of typically 500~meV as recently predicted\cite{Tawinan:2012a,Komsa:2012a} and experimentally confirmed \cite{He:2014a,Ugeda:2014a,Chernikov:2014a,Ye:2014a,Wang:2014a}. \\
\indent The physical properties drastically change when going from a ML to bilayers: First, in TMDC \textit{bilayers} the lowest energy optical transition is indirect in k-space, similar to the bulk material \cite{Mak:2010a,Splendiani:2010a,Jin:2013a,Zhang:2014a}. Time-integrated photoluminescence (PL) spectroscopy reveals in addition a higher energy direct transition associated to the recombination of carriers at the $K$ point. Second, crystal inversion symmetry is recovered in TMDC bilayers (the upper layer is rotated by 180$^\circ$ with respect to the lower one in 2$H$-WSe$_2$). As a result the chiral optical valley selectivity important for MLs should vanish, and no spin splitting in the bands is expected. Surprisingly, following polarized laser excitation, strongly polarized PL emission of the direct transition have been reported in time-integrated experiments recently in bilayer WSe$_2$ \cite{Jones:2014a} and WS$_2$ \cite{Zhu:2014b}. As possible origins, theoretical calculations suggest (i) intrinsic circular polarization in centrosymmetric even-layer stacks of TDMCs  \cite{Liu:2014a,Jones:2014a} and (ii) the excimer effect \cite{Yu:2014c}.\\
\begin{figure}[rh]
\includegraphics[width=0.39\textwidth]{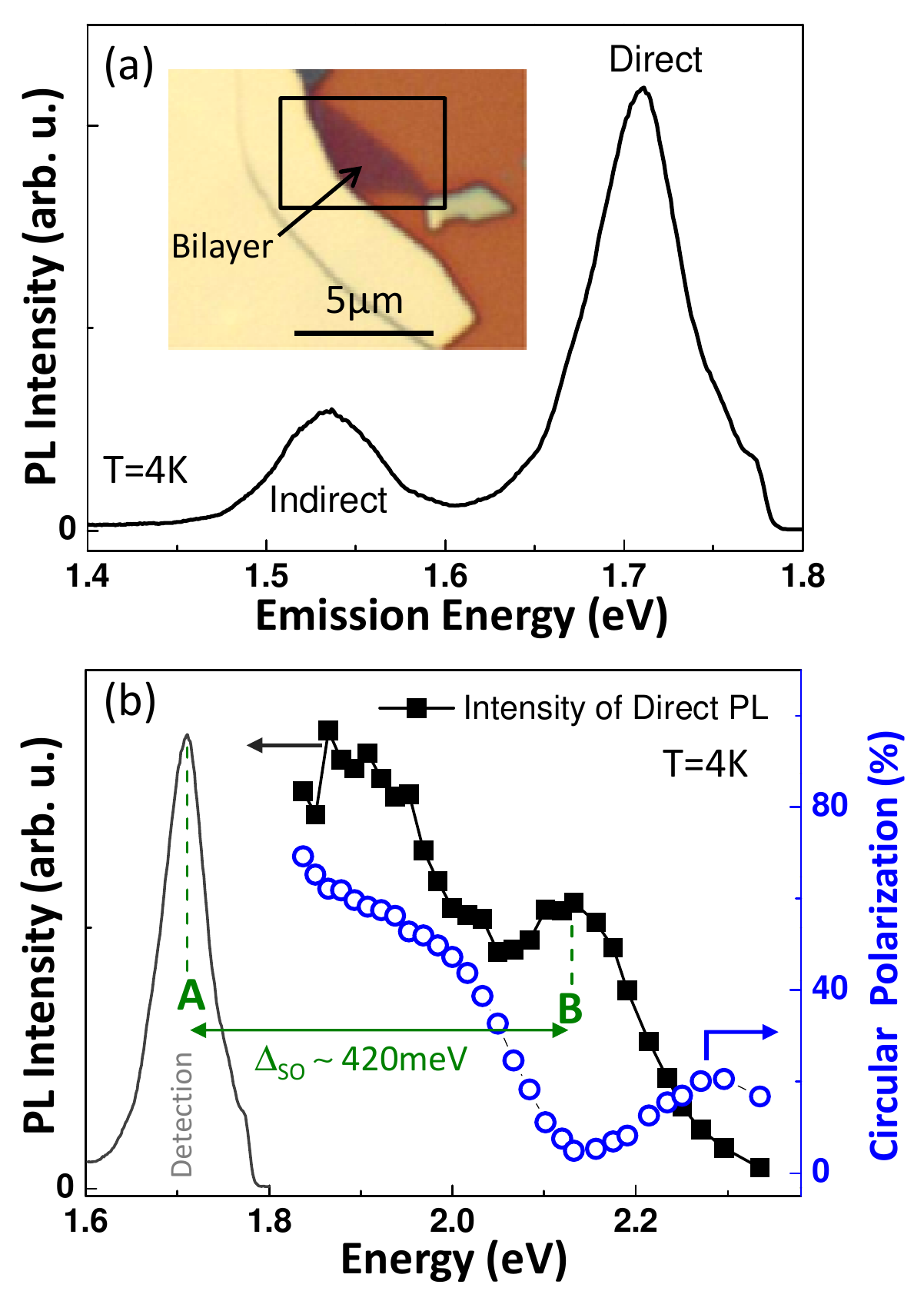}
\caption{\label{fig:fig1} (a) Time-integrated PL intensity of the WSe$_2$ bilayer exhibiting both Direct and Indirect optical transitions at T=4 K, $E_{\text{laser}}$~=1.851eV. Inset: Optical reflection image of a WSe$_2$ bilayer on Si substrate. (b) PL Excitation (PLE) spectroscopy: intensity of the Direct transition PL as a function of the excitation laser energy (black squares). Right axis: circular polarization $P_c$ direct transition PL (blue circles).
}
\end{figure} 
\indent To shed light on the competition between direct and indirect optical transitions and the intriguing spin and valley physics, we perform time- and polarization resolved PL experiments as a function of temperature and laser excitation energy.  In contrast to MLs, where the carrier dynamics has been measured by time-resolved absorption, reflection or PL spectroscopy \cite{Korn:2011a,Lagarde:2014a,Shi:2013b,Wang:2013d,Mai:2014a}, the measurement of both the direct and indirect optical transition kinetics have not been reported in TMDC bilayers to the best of our knowledge. We show that the direct transition in the bilayer is characterized by a very short decay time $\tau_D\lesssim3$~ps at low temperature, similarly to WSe$_2$ MLs \cite{Wang:2014b}. The exciton dynamics of the indirect transition is about ten times longer with a decay time of $\tau_I\approx25$~ps, with a weak temperature dependence. Moreover we demonstrate that the exciton (pseudo-)spin polarization and coherence measured on the direct optical transition occur on a very short time-scale of a few picoseconds. These results are also interesting in the context of TMDC heterostructures, proposed recently for photovoltaic or water-splitting applications where the knowledge of the electronic excitations dynamics is essential \cite{Rivera:2014a,Yu:2014d}.\\
\textit{Experimental set-up and sample.---} WSe$_2$ flakes are obtained by micro-mechanical cleavage of a bulk WSe$_2$ crystal (from 2D Semiconductors, USA) on 90 nm SiO$_2$ on a Si substrate. The bilayer region is identified by optical contrast (see inset of Fig.~\ref{fig:fig1}a) and very clearly in PL spectroscopy. Experiments between T=4 and 300K are carried out in a confocal microscope optimized for polarized PL experiments \cite{Urbaszek:2013a}. The WSe$_2$ bilayer is excited by picosecond pulses generated by a tunable frequency-doubled optical parametric oscillator (OPO) synchronously pumped by a mode-locked Ti:Sa laser. The typical pulse and spectral width are 1.6 ps and 3 meV respectively; the repetition rate is 80 MHz. The laser average power is in the 200 $\mu$W range, in the linear absorption regime. The detection spot diameter is $\approx1\mu$m. For time integrated experiments, the PL emission is dispersed in a spectrometer and detected with a Si-CCD camera. For time-resolved experiments, the PL signal is dispersed by an imaging spectrometer and detected by a synchro-scan Hamamatsu Streak Camera with an overall time resolution of 3 ps. The circular PL polarization $P_c$ is defined as $P_c=(I_{\sigma+}-I_{\sigma-})/(I_{\sigma+}+I_{\sigma-})$, where $I_{\sigma+}(I_{\sigma-})$denotes the intensity of the right ($\sigma+$) and left ($I_{\sigma-}$) circularly polarized emission. Similarly the linear PL polarization writes $P_l=(I_X-I_Y)/(I_X+I_Y)$ with $I_X(I_Y)$ the $X$ and $Y$ linearly polarized emission components.\\
\indent Time-integrated PL spectra at T=4 K for a laser excitation energy $E_{\text{laser}}$=1.851~eV are presented in Fig.~\ref{fig:fig1}a. Based on previous work \cite{Jones:2014a,Sahin:2013a,Zhao:2013b,Zhao:2013c}, the two transitions are attributed to the direct ($E$=1.711~eV) and indirect (1.535~eV) exciton radiative recombination in WSe$_2$ bilayers, respectively. The PL excitation (PLE) spectrum in Fig.\ref{fig:fig1}b for a detection energy set on the direct A-exciton transition ($E$=1.711~eV) exhibits a clear resonance for a laser energy $E_\text{laser}$=2.13~eV which corresponds to the excitation of the B-exciton. We find an SO-splitting energy between the A- and B-exciton of about 420 meV, in agreement with measurements in WSe$_2$ MLs \cite{Wang:2014b}.\\
\begin{figure}[rh]
\includegraphics[width=0.45\textwidth]{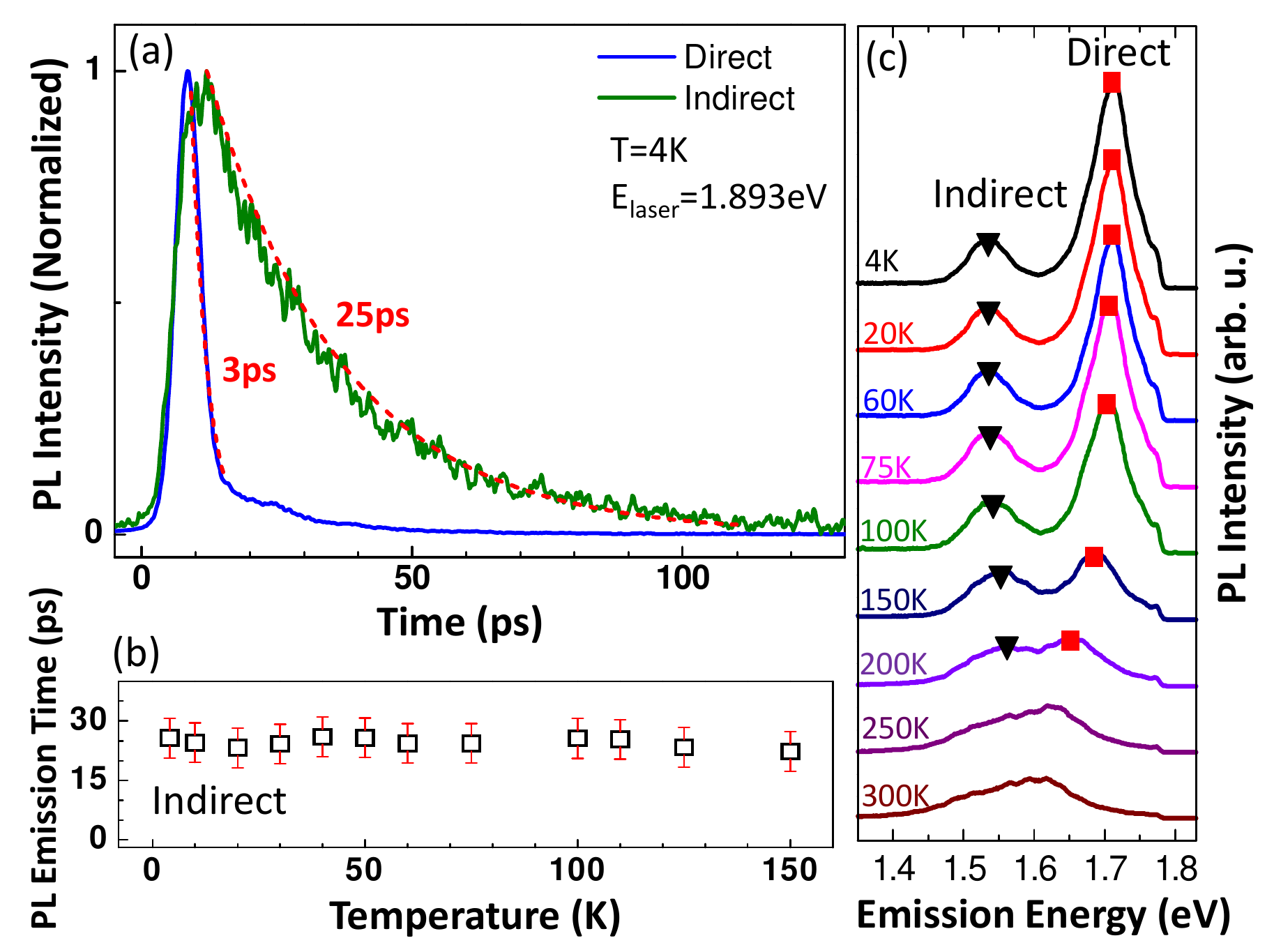}
\caption{\label{fig:fig2} (a) Normalized PL intensity as a function of time for the Indirect (green line) and Direct (blue line) transitions at $E$=1.535 eV and and $E$=1.711 eV respectively. The picosecond laser energy is $E_{\text{laser}}$=1.893 eV. The red dotted lines correspond to mono-exponential fits. (b) Temperature dependence of the PL emission time of the Indirect transition. (c) PL spectra of both the Direct and Indirect transitions as a function of temperature. The fitted Direct (Indirect) peak energy is indicated by red squares (black triangles).
}
\end{figure} 
\textit{Exciton dynamics and temperature dependence.---} In Fig.~\ref{fig:fig2}a the exciton kinetics for both optical transitions are displayed for T=4 K. Remarkably the indirect transition is characterized by a PL decay time ($\approx~25$~ps) about one order of magnitude longer than the direct one. We find a very fast recombination time for the direct transition ($\lesssim~3$~ps), as fast as the one measured for the neutral A-exciton in WSe$_2$ \cite{Wang:2014b} and MoS$_2$ MLs \cite{Lagarde:2014a,Korn:2011a}. This efficient and fast coupling to light of the direct transition explains the strong relative intensity of the direct transition in time-integrated PL compared to the indirect one observed in Fig.\ref{fig:fig1}a.\\
\indent Ab-initio calculations performed on WSe$_2$ bilayers predict that the direct transition corresponds to recombination of both electrons and holes lying at the extrema of the conduction band (CB) and the valence band (VB) in the K valley \cite{Zhao:2013c,Sahin:2013a,Debbichi:2014a}. The indirect transition is usually assigned to VB holes at the $K$ point and CB electrons lying in a minimum energy point located between $\Gamma$ and $K$ points \cite{type2}. As an indirect transitions requires in addition absorption or emission of a phonon \cite{Harley:1994a}, the measurement of a 10-times longer PL decay time for the lowest energy transition in WSe$_2$ bilayers compared to the one measured in MLs supports our interpretation. As far as band structure calculations are concerned, note however that the strong Coulomb effects undoubtedly present in TMDC bilayers (with exciton binding energies ~500 meV reported for MLs \cite{He:2014a,Ugeda:2014a,Chernikov:2014a,Ye:2014a,Wang:2014a}) are not taken into account for the optical transitions.\\
\begin{figure}[rh]
\includegraphics[width=0.495\textwidth]{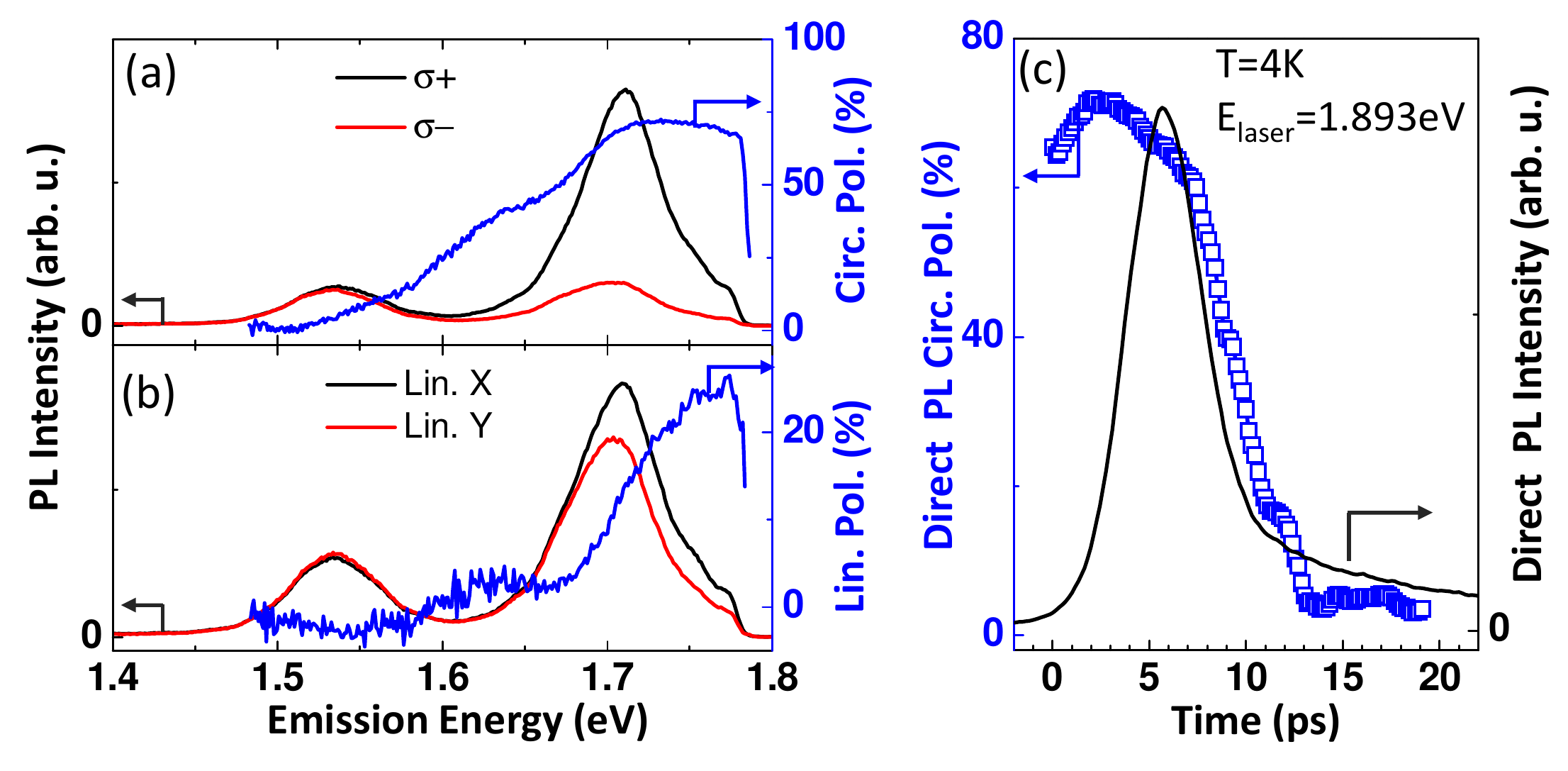}
\caption{\label{fig:fig3} T=4 K. (a) Laser polarization $\sigma^+$, $E_{\text{laser}}$=1.851 eV. Time-integrated PL spectra of bilayer WSe$_2$. Black (red) corresponds to $\sigma^+$($\sigma^-$) polarized emission. Right axis : circular PL polarization. (b) Laser polarized linear X. Black (red) corresponds to linear X (linear Y) polarized emission. Right axis: PL linear polarization. (c) T=4 K, $E_{\text{laser}}$=1.893 eV, PL circular polarization dynamics (blue squares) for the Direct transition ($E$= 1.711 eV). Right axis: PL intensity dynamics.
}
\end{figure} 
\indent The temperature dependence of the direct and indirect transition PL spectra is displayed in Fig.~\ref{fig:fig2}c. Both transitions can be observed up to room temperature though there is a significant overlap in energy above 150 K. The clear observation of the direct transition for all temperatures demonstrates that the coupling to light (with a strong oscillator strength) of high energy carriers is comparable to the energy relaxation time down to k-valley point where the indirect transition occurs. We observe in Fig.\ref{fig:fig2}b that this indirect transition is characterized by a decay time of  about 25 ps, independent of the temperature  in the range 4 to 150 K. For higher temperatures the lifetime determination is ambiguous due to the energy overlap of direct and indirect transitions.\\
\indent \textit{PL polarization dynamics.---} TMDC bilayers have unique polarization properties \cite{Jones:2014a,Zhu:2014b,Liu:2014a,Yu:2014c}. They are part of an interesting class of materials that consist of individual, inversion-asymmetric layers with strong SO-coupling that from a globally symmetric stack \cite{Zhang:2014b,Riley:2014a}.  Fig.~\ref{fig:fig3}a presents the circularly polarized PL components ($\sigma^+$ and $\sigma^-$) following $\sigma^+$ polarized laser excitation. We observe a large circular polarization of about 70\% of the direct transition whereas no polarization is observed on the indirect transition. We record in Fig.~\ref{fig:fig1}b a decrease of the circular PL  polarisation of the direct transition as the laser energy increases, similar to the observations for MoS$_2$ and WSe$_2$ monolayers \cite{Lagarde:2014a,Kioseoglou:2012a,Wang:2014a}. The measured $P_c$ is roughly constant across the direct exciton emission spectrum in Fig.~\ref{fig:fig3}a, that contains contributions from neutral excitons on the high energy side and charged excitons (trions) on the low energy side.\\
\indent In addition we record a significant linear PL polarization of up to 25\% of the direct transition following linearly polarized laser excitation \cite{imppol} as a result of the coherent superposition of exciton states as shown in Fig.~\ref{fig:fig3}b \cite{Jones:2014a}. The linear PL polarization is always aligned with the in-plane laser polarization and not a particular crystallographic axis. The maximum measured $P_l$ in Fig.~\ref{fig:fig3}b is observed at the high energy side of the PL peak at $E\approx 1.74$~eV, which coincides exactly with the neutral exciton energy identified by \textcite{Jones:2014a}. We tentatively attribute the low $P_l$ at the low energy side of the direct emission to trions.  \\
\indent Initially, chiral optial selection rules and hence strong PL polarization for excitonic transitions were only predicted for TMDC MLs \cite{Xiao:2012a}, as inversion symmetry is globally restored for 2$H$-bilayers. In that respect the large values of $P_c$ and $P_l$ in Fig.~\ref{fig:fig3} are remarkable. The high $P_c$ and $P_l$ in \textit{time-integrated} PL measurements of the direct transition of neutral or charged excitons in WSe$_2$ \cite{Jones:2014a} and WS$_2$  \cite{Zhu:2014b} bilayers were interpreted recently as a consequence of the enhancement of the exciton spin lifetime. This is thought to be due to a spin-layer locking effect \cite{Jones:2014a}, where the exciton polarization is locked to the layer index i.e. its localization on the upper or lower layer. This is a fascinating subject for theory due the competition between the strong SO coupling (measured to be 420~meV in our sample in Fig.~\ref{fig:fig1}b), layer hopping energies and Coulomb effects \cite{Jones:2014a,Liu:2014a,Yu:2014c,Zhang:2014b}. Here \textit{time-resolved} PL experiments are crucial and we demonstrate in Fig.~\ref{fig:fig3}c that $P_c$ of the direct transition decays within a few picoseconds, as fast as the PL intensity which is shown for comparison. This fast polarization decay could be due the large exciton exchange exchange interaction for excitons in bilayers (direct transition) similarly to the exciton depolarization evidenced recently in individual TMDC MLs \cite{Glazov:2014a,Yu:2014a,Zhucr:2014a}. In addition to polarization relaxation processes present already in isolated MLs, new depolarization channels open up in bilayers due to interlayer coupling \cite{Jones:2014a,Liu:2014a}. Interestingly in Fig.~\ref{fig:fig1}b we observe a global minimum of $P_c$ of the A-exciton when the B-exciton is resonantly generated. One could speculate that at this energy interlayer hopping and spin flips, blocked by the SO interaction at lower energies, become more likely. Note that we do not observe in Fig.~\ref{fig:fig3}c the extremely long PL polarization decay time which should be the fingerprint of an excimer transition proposed by Yu et al.\cite{Yu:2014c}.\\
\indent In conclusion, we have measured the exciton dynamics for both direct and indirect optical transitions in WSe$_2$ bilayers. These results reveal fast recombination times ($\lesssim 3$~ps and $\approx 25$~ps respectively) which should be taken into account to explain the unique spin polarization properties of the TMDC bilayers as well as the various proposed applications of such nanostructures including high temperature superfluidity based on indirect excitons \cite{Fogler:2014a}.\\
\indent We acknowledge partial funding from ERC Grant No. 306719 and Programme Investissements d'Avenir ANR-11-IDEX-0002-02, reference ANR-10-LABX-0037- NEXT.


\begin{thebibliography}{46}
\expandafter\ifx\csname natexlab\endcsname\relax\def\natexlab#1{#1}\fi
\expandafter\ifx\csname bibnamefont\endcsname\relax
  \def\bibnamefont#1{#1}\fi
\expandafter\ifx\csname bibfnamefont\endcsname\relax
  \def\bibfnamefont#1{#1}\fi
\expandafter\ifx\csname citenamefont\endcsname\relax
  \def\citenamefont#1{#1}\fi
\expandafter\ifx\csname url\endcsname\relax
  \def\url#1{\texttt{#1}}\fi
\expandafter\ifx\csname urlprefix\endcsname\relax\def\urlprefix{URL }\fi
\providecommand{\bibinfo}[2]{#2}
\providecommand{\eprint}[2][]{\url{#2}}

\bibitem[{\citenamefont{Geim and Grigorieva}(2013)}]{Geim:2013a}
\bibinfo{author}{\bibfnamefont{A.~K.} \bibnamefont{Geim}} \bibnamefont{and}
  \bibinfo{author}{\bibfnamefont{I.~V.} \bibnamefont{Grigorieva}},
  \bibinfo{journal}{Nature} \textbf{\bibinfo{volume}{499}},
  \bibinfo{pages}{419–} (\bibinfo{year}{2013}).

\bibitem[{\citenamefont{Xu et~al.}(2014)\citenamefont{Xu, Xiao, Heinz, and
  Yao}}]{Xu:2014a}
\bibinfo{author}{\bibfnamefont{X.}~\bibnamefont{Xu}},
  \bibinfo{author}{\bibfnamefont{D.}~\bibnamefont{Xiao}},
  \bibinfo{author}{\bibfnamefont{T.~F.} \bibnamefont{Heinz}}, \bibnamefont{and}
  \bibinfo{author}{\bibfnamefont{W.}~\bibnamefont{Yao}},
  \bibinfo{journal}{Nature Physics} \textbf{\bibinfo{volume}{10}},
  \bibinfo{pages}{343} (\bibinfo{year}{2014}).

\bibitem[{\citenamefont{Mak et~al.}(2010)\citenamefont{Mak, Lee, Hone, Shan,
  and Heinz}}]{Mak:2010a}
\bibinfo{author}{\bibfnamefont{K.~F.} \bibnamefont{Mak}},
  \bibinfo{author}{\bibfnamefont{C.}~\bibnamefont{Lee}},
  \bibinfo{author}{\bibfnamefont{J.}~\bibnamefont{Hone}},
  \bibinfo{author}{\bibfnamefont{J.}~\bibnamefont{Shan}}, \bibnamefont{and}
  \bibinfo{author}{\bibfnamefont{T.~F.} \bibnamefont{Heinz}},
  \bibinfo{journal}{Phys. Rev. Lett.} \textbf{\bibinfo{volume}{105}},
  \bibinfo{pages}{136805} (\bibinfo{year}{2010}).

\bibitem[{\citenamefont{Splendiani et~al.}(2010)\citenamefont{Splendiani, Sun,
  Zhang, Li, Kim, Chim, Galli, and Wang}}]{Splendiani:2010a}
\bibinfo{author}{\bibfnamefont{A.}~\bibnamefont{Splendiani}},
  \bibinfo{author}{\bibfnamefont{L.}~\bibnamefont{Sun}},
  \bibinfo{author}{\bibfnamefont{Y.}~\bibnamefont{Zhang}},
  \bibinfo{author}{\bibfnamefont{T.}~\bibnamefont{Li}},
  \bibinfo{author}{\bibfnamefont{J.}~\bibnamefont{Kim}},
  \bibinfo{author}{\bibfnamefont{C.-Y.} \bibnamefont{Chim}},
  \bibinfo{author}{\bibfnamefont{G.}~\bibnamefont{Galli}}, \bibnamefont{and}
  \bibinfo{author}{\bibfnamefont{F.}~\bibnamefont{Wang}},
  \bibinfo{journal}{Nano Letters} \textbf{\bibinfo{volume}{10}},
  \bibinfo{pages}{1271} (\bibinfo{year}{2010}).

\bibitem[{\citenamefont{Xiao et~al.}(2012)\citenamefont{Xiao, Liu, Feng, Xu,
  and Yao}}]{Xiao:2012a}
\bibinfo{author}{\bibfnamefont{D.}~\bibnamefont{Xiao}},
  \bibinfo{author}{\bibfnamefont{G.-B.} \bibnamefont{Liu}},
  \bibinfo{author}{\bibfnamefont{W.}~\bibnamefont{Feng}},
  \bibinfo{author}{\bibfnamefont{X.}~\bibnamefont{Xu}}, \bibnamefont{and}
  \bibinfo{author}{\bibfnamefont{W.}~\bibnamefont{Yao}},
  \bibinfo{journal}{Phys. Rev. Lett.} \textbf{\bibinfo{volume}{108}},
  \bibinfo{pages}{196802} (\bibinfo{year}{2012}).

\bibitem[{\citenamefont{Cao et~al.}(2012)\citenamefont{Cao, Wang, Han, Ye, Zhu,
  Shi, Niu, Tan, Wang, Liu et~al.}}]{Cao:2012a}
\bibinfo{author}{\bibfnamefont{T.}~\bibnamefont{Cao}},
  \bibinfo{author}{\bibfnamefont{G.}~\bibnamefont{Wang}},
  \bibinfo{author}{\bibfnamefont{W.}~\bibnamefont{Han}},
  \bibinfo{author}{\bibfnamefont{H.}~\bibnamefont{Ye}},
  \bibinfo{author}{\bibfnamefont{C.}~\bibnamefont{Zhu}},
  \bibinfo{author}{\bibfnamefont{J.}~\bibnamefont{Shi}},
  \bibinfo{author}{\bibfnamefont{Q.}~\bibnamefont{Niu}},
  \bibinfo{author}{\bibfnamefont{P.}~\bibnamefont{Tan}},
  \bibinfo{author}{\bibfnamefont{E.}~\bibnamefont{Wang}},
  \bibinfo{author}{\bibfnamefont{B.}~\bibnamefont{Liu}}, \bibnamefont{et~al.},
  \bibinfo{journal}{Nature Communications} \textbf{\bibinfo{volume}{3}},
  \bibinfo{pages}{887} (\bibinfo{year}{2012}).

\bibitem[{\citenamefont{Mak et~al.}(2012)\citenamefont{Mak, He, Shan, and
  Heinz}}]{Mak:2012a}
\bibinfo{author}{\bibfnamefont{K.~F.} \bibnamefont{Mak}},
  \bibinfo{author}{\bibfnamefont{K.}~\bibnamefont{He}},
  \bibinfo{author}{\bibfnamefont{J.}~\bibnamefont{Shan}}, \bibnamefont{and}
  \bibinfo{author}{\bibfnamefont{T.~F.} \bibnamefont{Heinz}},
  \bibinfo{journal}{Nat. Nanotechnol.} \textbf{\bibinfo{volume}{7}},
  \bibinfo{pages}{494} (\bibinfo{year}{2012}).

\bibitem[{\citenamefont{Sallen et~al.}(2012)\citenamefont{Sallen, Bouet, Marie,
  Wang, Zhu, Han, Lu, Tan, Amand, Liu et~al.}}]{Sallen:2012a}
\bibinfo{author}{\bibfnamefont{G.}~\bibnamefont{Sallen}},
  \bibinfo{author}{\bibfnamefont{L.}~\bibnamefont{Bouet}},
  \bibinfo{author}{\bibfnamefont{X.}~\bibnamefont{Marie}},
  \bibinfo{author}{\bibfnamefont{G.}~\bibnamefont{Wang}},
  \bibinfo{author}{\bibfnamefont{C.~R.} \bibnamefont{Zhu}},
  \bibinfo{author}{\bibfnamefont{W.~P.} \bibnamefont{Han}},
  \bibinfo{author}{\bibfnamefont{Y.}~\bibnamefont{Lu}},
  \bibinfo{author}{\bibfnamefont{P.~H.} \bibnamefont{Tan}},
  \bibinfo{author}{\bibfnamefont{T.}~\bibnamefont{Amand}},
  \bibinfo{author}{\bibfnamefont{B.~L.} \bibnamefont{Liu}},
  \bibnamefont{et~al.}, \bibinfo{journal}{Phys. Rev. B}
  \textbf{\bibinfo{volume}{86}}, \bibinfo{pages}{081301}
  (\bibinfo{year}{2012}).

\bibitem[{\citenamefont{Kioseoglou et~al.}(2012)\citenamefont{Kioseoglou,
  Hanbicki, Currie, Friedman, Gunlycke, and Jonker}}]{Kioseoglou:2012a}
\bibinfo{author}{\bibfnamefont{G.}~\bibnamefont{Kioseoglou}},
  \bibinfo{author}{\bibfnamefont{A.~T.} \bibnamefont{Hanbicki}},
  \bibinfo{author}{\bibfnamefont{M.}~\bibnamefont{Currie}},
  \bibinfo{author}{\bibfnamefont{A.~L.} \bibnamefont{Friedman}},
  \bibinfo{author}{\bibfnamefont{D.}~\bibnamefont{Gunlycke}}, \bibnamefont{and}
  \bibinfo{author}{\bibfnamefont{B.~T.} \bibnamefont{Jonker}},
  \bibinfo{journal}{Applied Physics Letters} \textbf{\bibinfo{volume}{101}},
  \bibinfo{eid}{221907} (pages~\bibinfo{numpages}{4}) (\bibinfo{year}{2012}).

\bibitem[{\citenamefont{Jones et~al.}(2013)\citenamefont{Jones, Yu, Ghimire,
  Wu, Aivazian, Ross, Zhao, Yan, Mandrus, Xiao et~al.}}]{Jones:2013a}
\bibinfo{author}{\bibfnamefont{A.~M.} \bibnamefont{Jones}},
  \bibinfo{author}{\bibfnamefont{H.}~\bibnamefont{Yu}},
  \bibinfo{author}{\bibfnamefont{N.~J.} \bibnamefont{Ghimire}},
  \bibinfo{author}{\bibfnamefont{S.}~\bibnamefont{Wu}},
  \bibinfo{author}{\bibfnamefont{G.}~\bibnamefont{Aivazian}},
  \bibinfo{author}{\bibfnamefont{J.~S.} \bibnamefont{Ross}},
  \bibinfo{author}{\bibfnamefont{B.}~\bibnamefont{Zhao}},
  \bibinfo{author}{\bibfnamefont{J.}~\bibnamefont{Yan}},
  \bibinfo{author}{\bibfnamefont{D.~G.} \bibnamefont{Mandrus}},
  \bibinfo{author}{\bibfnamefont{D.}~\bibnamefont{Xiao}}, \bibnamefont{et~al.},
  \bibinfo{journal}{Nat. Nanotechnol.} \textbf{\bibinfo{volume}{8}},
  \bibinfo{pages}{634} (\bibinfo{year}{2013}).

\bibitem[{\citenamefont{Mak et~al.}(2014)\citenamefont{Mak, McGill, Park, and
  McEuen}}]{Mak:2014a}
\bibinfo{author}{\bibfnamefont{K.~F.} \bibnamefont{Mak}},
  \bibinfo{author}{\bibfnamefont{K.~L.} \bibnamefont{McGill}},
  \bibinfo{author}{\bibfnamefont{J.}~\bibnamefont{Park}}, \bibnamefont{and}
  \bibinfo{author}{\bibfnamefont{P.~L.} \bibnamefont{McEuen}},
  \bibinfo{journal}{Science} \textbf{\bibinfo{volume}{344}},
  \bibinfo{pages}{1489} (\bibinfo{year}{2014}).

\bibitem[{\citenamefont{Cheiwchanchamnangij and
  Lambrecht}(2012)}]{Tawinan:2012a}
\bibinfo{author}{\bibfnamefont{T.}~\bibnamefont{Cheiwchanchamnangij}}
  \bibnamefont{and} \bibinfo{author}{\bibfnamefont{W.~R.~L.}
  \bibnamefont{Lambrecht}}, \bibinfo{journal}{Phys. Rev. B}
  \textbf{\bibinfo{volume}{85}}, \bibinfo{pages}{205302}
  (\bibinfo{year}{2012}).

\bibitem[{\citenamefont{Komsa and Krasheninnikov}(2012)}]{Komsa:2012a}
\bibinfo{author}{\bibfnamefont{H.-P.} \bibnamefont{Komsa}} \bibnamefont{and}
  \bibinfo{author}{\bibfnamefont{A.~V.} \bibnamefont{Krasheninnikov}},
  \bibinfo{journal}{Phys. Rev. B} \textbf{\bibinfo{volume}{86}},
  \bibinfo{pages}{241201} (\bibinfo{year}{2012}).

\bibitem[{\citenamefont{He et~al.}(2014)\citenamefont{He, Kumar, Zhao, Wang,
  Mak, Zhao, and Shan}}]{He:2014a}
\bibinfo{author}{\bibfnamefont{K.}~\bibnamefont{He}},
  \bibinfo{author}{\bibfnamefont{N.}~\bibnamefont{Kumar}},
  \bibinfo{author}{\bibfnamefont{L.}~\bibnamefont{Zhao}},
  \bibinfo{author}{\bibfnamefont{Z.}~\bibnamefont{Wang}},
  \bibinfo{author}{\bibfnamefont{K.~F.} \bibnamefont{Mak}},
  \bibinfo{author}{\bibfnamefont{H.}~\bibnamefont{Zhao}}, \bibnamefont{and}
  \bibinfo{author}{\bibfnamefont{J.}~\bibnamefont{Shan}},
  \bibinfo{journal}{Phys. Rev. Lett.} \textbf{\bibinfo{volume}{113}},
  \bibinfo{pages}{026803} (\bibinfo{year}{2014}).

\bibitem[{\citenamefont{{Ugeda} et~al.}(2014)\citenamefont{{Ugeda}, {Bradley},
  {Shi}, {da Jornada}, {Zhang}, {Qiu}, {Mo}, {Hussain}, {Shen}, {Wang}
  et~al.}}]{Ugeda:2014a}
\bibinfo{author}{\bibfnamefont{M.~M.} \bibnamefont{{Ugeda}}},
  \bibinfo{author}{\bibfnamefont{A.~J.} \bibnamefont{{Bradley}}},
  \bibinfo{author}{\bibfnamefont{S.-F.} \bibnamefont{{Shi}}},
  \bibinfo{author}{\bibfnamefont{F.~H.} \bibnamefont{{da Jornada}}},
  \bibinfo{author}{\bibfnamefont{Y.}~\bibnamefont{{Zhang}}},
  \bibinfo{author}{\bibfnamefont{D.~Y.} \bibnamefont{{Qiu}}},
  \bibinfo{author}{\bibfnamefont{S.-K.} \bibnamefont{{Mo}}},
  \bibinfo{author}{\bibfnamefont{Z.}~\bibnamefont{{Hussain}}},
  \bibinfo{author}{\bibfnamefont{Z.-X.} \bibnamefont{{Shen}}},
  \bibinfo{author}{\bibfnamefont{F.}~\bibnamefont{{Wang}}},
  \bibnamefont{et~al.}, \bibinfo{journal}{Nature Materials}
  \textbf{\bibinfo{volume}{doi: 10.1038/nmat4061}} (\bibinfo{year}{2014}).

\bibitem[{\citenamefont{Chernikov et~al.}(2014)\citenamefont{Chernikov,
  Berkelbach, Hill, Rigosi, Li, Aslan, Reichman, Hybertsen, and
  Heinz}}]{Chernikov:2014a}
\bibinfo{author}{\bibfnamefont{A.}~\bibnamefont{Chernikov}},
  \bibinfo{author}{\bibfnamefont{T.~C.} \bibnamefont{Berkelbach}},
  \bibinfo{author}{\bibfnamefont{H.~M.} \bibnamefont{Hill}},
  \bibinfo{author}{\bibfnamefont{A.}~\bibnamefont{Rigosi}},
  \bibinfo{author}{\bibfnamefont{Y.}~\bibnamefont{Li}},
  \bibinfo{author}{\bibfnamefont{O.~B.} \bibnamefont{Aslan}},
  \bibinfo{author}{\bibfnamefont{D.~R.} \bibnamefont{Reichman}},
  \bibinfo{author}{\bibfnamefont{M.~S.} \bibnamefont{Hybertsen}},
  \bibnamefont{and} \bibinfo{author}{\bibfnamefont{T.~F.} \bibnamefont{Heinz}},
  \bibinfo{journal}{Phys. Rev. Lett.} \textbf{\bibinfo{volume}{113}},
  \bibinfo{pages}{076802} (\bibinfo{year}{2014}).

\bibitem[{\citenamefont{{Ye} et~al.}(2014)\citenamefont{{Ye}, {Cao}, {O'Brien},
  {Zhu}, {Yin}, {Wang}, {Louie}, and {Zhang}}}]{Ye:2014a}
\bibinfo{author}{\bibfnamefont{Z.}~\bibnamefont{{Ye}}},
  \bibinfo{author}{\bibfnamefont{T.}~\bibnamefont{{Cao}}},
  \bibinfo{author}{\bibfnamefont{K.}~\bibnamefont{{O'Brien}}},
  \bibinfo{author}{\bibfnamefont{H.}~\bibnamefont{{Zhu}}},
  \bibinfo{author}{\bibfnamefont{X.}~\bibnamefont{{Yin}}},
  \bibinfo{author}{\bibfnamefont{Y.}~\bibnamefont{{Wang}}},
  \bibinfo{author}{\bibfnamefont{S.~G.} \bibnamefont{{Louie}}},
  \bibnamefont{and} \bibinfo{author}{\bibfnamefont{X.}~\bibnamefont{{Zhang}}},
  \bibinfo{journal}{Nature} \textbf{\bibinfo{volume}{513}},
  \bibinfo{pages}{214} (\bibinfo{year}{2014}).

\bibitem[{\citenamefont{Wang et~al.}(2014{\natexlab{a}})\citenamefont{Wang,
  Marie, Gerber, Amand, Lagarde, Bouet, Vidal, Balocchi, and
  Urbaszek}}]{Wang:2014a}
\bibinfo{author}{\bibfnamefont{G.}~\bibnamefont{Wang}},
  \bibinfo{author}{\bibfnamefont{X.}~\bibnamefont{Marie}},
  \bibinfo{author}{\bibfnamefont{I.}~\bibnamefont{Gerber}},
  \bibinfo{author}{\bibfnamefont{T.}~\bibnamefont{Amand}},
  \bibinfo{author}{\bibfnamefont{D.}~\bibnamefont{Lagarde}},
  \bibinfo{author}{\bibfnamefont{L.}~\bibnamefont{Bouet}},
  \bibinfo{author}{\bibfnamefont{M.}~\bibnamefont{Vidal}},
  \bibinfo{author}{\bibfnamefont{A.}~\bibnamefont{Balocchi}}, \bibnamefont{and}
  \bibinfo{author}{\bibfnamefont{B.}~\bibnamefont{Urbaszek}},
  \bibinfo{journal}{e-print} \textbf{\bibinfo{volume}{arXiv:1404.0056}}
  (\bibinfo{year}{2014}{\natexlab{a}}).

\bibitem[{\citenamefont{Jin et~al.}(2013)\citenamefont{Jin, Yeh, Zaki, Zhang,
  Sadowski, Al-Mahboob, van~der Zande, Chenet, Dadap, Herman
  et~al.}}]{Jin:2013a}
\bibinfo{author}{\bibfnamefont{W.}~\bibnamefont{Jin}},
  \bibinfo{author}{\bibfnamefont{P.-C.} \bibnamefont{Yeh}},
  \bibinfo{author}{\bibfnamefont{N.}~\bibnamefont{Zaki}},
  \bibinfo{author}{\bibfnamefont{D.}~\bibnamefont{Zhang}},
  \bibinfo{author}{\bibfnamefont{J.~T.} \bibnamefont{Sadowski}},
  \bibinfo{author}{\bibfnamefont{A.}~\bibnamefont{Al-Mahboob}},
  \bibinfo{author}{\bibfnamefont{A.~M.} \bibnamefont{van~der Zande}},
  \bibinfo{author}{\bibfnamefont{D.~A.} \bibnamefont{Chenet}},
  \bibinfo{author}{\bibfnamefont{J.~I.} \bibnamefont{Dadap}},
  \bibinfo{author}{\bibfnamefont{I.~P.} \bibnamefont{Herman}},
  \bibnamefont{et~al.}, \bibinfo{journal}{Phys. Rev. Lett.}
  \textbf{\bibinfo{volume}{111}}, \bibinfo{pages}{106801}
  (\bibinfo{year}{2013}).

\bibitem[{\citenamefont{Zhang et~al.}(2014{\natexlab{a}})\citenamefont{Zhang,
  Chang, Zhou, Cui, Yan, Liu, Schmitt, Lee, Moore, Chen et~al.}}]{Zhang:2014a}
\bibinfo{author}{\bibfnamefont{Y.}~\bibnamefont{Zhang}},
  \bibinfo{author}{\bibfnamefont{T.-R.} \bibnamefont{Chang}},
  \bibinfo{author}{\bibfnamefont{B.}~\bibnamefont{Zhou}},
  \bibinfo{author}{\bibfnamefont{Y.-T.} \bibnamefont{Cui}},
  \bibinfo{author}{\bibfnamefont{H.}~\bibnamefont{Yan}},
  \bibinfo{author}{\bibfnamefont{Z.}~\bibnamefont{Liu}},
  \bibinfo{author}{\bibfnamefont{F.}~\bibnamefont{Schmitt}},
  \bibinfo{author}{\bibfnamefont{J.}~\bibnamefont{Lee}},
  \bibinfo{author}{\bibfnamefont{R.}~\bibnamefont{Moore}},
  \bibinfo{author}{\bibfnamefont{Y.}~\bibnamefont{Chen}}, \bibnamefont{et~al.},
  \bibinfo{journal}{Nature Nanotechnology} \textbf{\bibinfo{volume}{9}},
  \bibinfo{pages}{111} (\bibinfo{year}{2014}{\natexlab{a}}).

\bibitem[{\citenamefont{Jones et~al.}(2014)\citenamefont{Jones, Yu, Ross,
  Klement, Ghimire, Yan, Mandrus, Yao, and Xu}}]{Jones:2014a}
\bibinfo{author}{\bibfnamefont{A.~M.} \bibnamefont{Jones}},
  \bibinfo{author}{\bibfnamefont{H.}~\bibnamefont{Yu}},
  \bibinfo{author}{\bibfnamefont{J.~S.} \bibnamefont{Ross}},
  \bibinfo{author}{\bibfnamefont{P.}~\bibnamefont{Klement}},
  \bibinfo{author}{\bibfnamefont{N.~J.} \bibnamefont{Ghimire}},
  \bibinfo{author}{\bibfnamefont{J.}~\bibnamefont{Yan}},
  \bibinfo{author}{\bibfnamefont{D.~G.} \bibnamefont{Mandrus}},
  \bibinfo{author}{\bibfnamefont{W.}~\bibnamefont{Yao}}, \bibnamefont{and}
  \bibinfo{author}{\bibfnamefont{X.}~\bibnamefont{Xu}}, \bibinfo{journal}{Nat.
  Phys} \textbf{\bibinfo{volume}{10}}, \bibinfo{pages}{130}
  (\bibinfo{year}{2014}).

\bibitem[{\citenamefont{Zhu et~al.}(2014)\citenamefont{Zhu, Zeng, Dai, Gong,
  and Cui}}]{Zhu:2014b}
\bibinfo{author}{\bibfnamefont{B.}~\bibnamefont{Zhu}},
  \bibinfo{author}{\bibfnamefont{H.}~\bibnamefont{Zeng}},
  \bibinfo{author}{\bibfnamefont{J.}~\bibnamefont{Dai}},
  \bibinfo{author}{\bibfnamefont{Z.}~\bibnamefont{Gong}}, \bibnamefont{and}
  \bibinfo{author}{\bibfnamefont{X.}~\bibnamefont{Cui}},
  \bibinfo{journal}{Proceedings of the National Academy of Sciences}
  \textbf{\bibinfo{volume}{111}}, \bibinfo{pages}{11606}
  (\bibinfo{year}{2014}).

\bibitem[{\citenamefont{{Liu} et~al.}(2014)\citenamefont{{Liu}, {Zhang}, and
  {Zunger}}}]{Liu:2014a}
\bibinfo{author}{\bibfnamefont{Q.}~\bibnamefont{{Liu}}},
  \bibinfo{author}{\bibfnamefont{X.}~\bibnamefont{{Zhang}}}, \bibnamefont{and}
  \bibinfo{author}{\bibfnamefont{A.}~\bibnamefont{{Zunger}}},
  \bibinfo{journal}{ArXiv e-prints}  (\bibinfo{year}{2014}),
  \eprint{1408.6001}.

\bibitem[{\citenamefont{Yu and Wu}(2014{\natexlab{a}})}]{Yu:2014c}
\bibinfo{author}{\bibfnamefont{T.}~\bibnamefont{Yu}} \bibnamefont{and}
  \bibinfo{author}{\bibfnamefont{M.~W.} \bibnamefont{Wu}},
  \bibinfo{journal}{Phys. Rev. B} \textbf{\bibinfo{volume}{90}},
  \bibinfo{pages}{035437} (\bibinfo{year}{2014}{\natexlab{a}}).

\bibitem[{\citenamefont{Korn et~al.}(2011)\citenamefont{Korn, Heydrich, Hirmer,
  Schmutzler, and Sch\"{u}ller}}]{Korn:2011a}
\bibinfo{author}{\bibfnamefont{T.}~\bibnamefont{Korn}},
  \bibinfo{author}{\bibfnamefont{S.}~\bibnamefont{Heydrich}},
  \bibinfo{author}{\bibfnamefont{M.}~\bibnamefont{Hirmer}},
  \bibinfo{author}{\bibfnamefont{J.}~\bibnamefont{Schmutzler}},
  \bibnamefont{and}
  \bibinfo{author}{\bibfnamefont{C.}~\bibnamefont{Sch\"{u}ller}},
  \bibinfo{journal}{Applied Physics Letters} \textbf{\bibinfo{volume}{99}},
  \bibinfo{eid}{102109} (\bibinfo{year}{2011}).

\bibitem[{\citenamefont{Lagarde et~al.}(2014)\citenamefont{Lagarde, Bouet,
  Marie, Zhu, Liu, Amand, Tan, and Urbaszek}}]{Lagarde:2014a}
\bibinfo{author}{\bibfnamefont{D.}~\bibnamefont{Lagarde}},
  \bibinfo{author}{\bibfnamefont{L.}~\bibnamefont{Bouet}},
  \bibinfo{author}{\bibfnamefont{X.}~\bibnamefont{Marie}},
  \bibinfo{author}{\bibfnamefont{C.~R.} \bibnamefont{Zhu}},
  \bibinfo{author}{\bibfnamefont{B.~L.} \bibnamefont{Liu}},
  \bibinfo{author}{\bibfnamefont{T.}~\bibnamefont{Amand}},
  \bibinfo{author}{\bibfnamefont{P.~H.} \bibnamefont{Tan}}, \bibnamefont{and}
  \bibinfo{author}{\bibfnamefont{B.}~\bibnamefont{Urbaszek}},
  \bibinfo{journal}{Phys. Rev. Lett.} \textbf{\bibinfo{volume}{112}},
  \bibinfo{pages}{047401} (\bibinfo{year}{2014}).

\bibitem[{\citenamefont{Shi et~al.}(2013)\citenamefont{Shi, Yan, Bertolazzi,
  Brivio, Gao, Kis, Jena, Xing, and Huang}}]{Shi:2013b}
\bibinfo{author}{\bibfnamefont{H.}~\bibnamefont{Shi}},
  \bibinfo{author}{\bibfnamefont{R.}~\bibnamefont{Yan}},
  \bibinfo{author}{\bibfnamefont{S.}~\bibnamefont{Bertolazzi}},
  \bibinfo{author}{\bibfnamefont{J.}~\bibnamefont{Brivio}},
  \bibinfo{author}{\bibfnamefont{B.}~\bibnamefont{Gao}},
  \bibinfo{author}{\bibfnamefont{A.}~\bibnamefont{Kis}},
  \bibinfo{author}{\bibfnamefont{D.}~\bibnamefont{Jena}},
  \bibinfo{author}{\bibfnamefont{H.~G.} \bibnamefont{Xing}}, \bibnamefont{and}
  \bibinfo{author}{\bibfnamefont{L.}~\bibnamefont{Huang}},
  \bibinfo{journal}{ACS Nano} \textbf{\bibinfo{volume}{7}},
  \bibinfo{pages}{1072} (\bibinfo{year}{2013}).

\bibitem[{\citenamefont{Wang et~al.}(2013)\citenamefont{Wang, Ge, Li, Qiu, Ji,
  Feng, and Sun}}]{Wang:2013d}
\bibinfo{author}{\bibfnamefont{Q.}~\bibnamefont{Wang}},
  \bibinfo{author}{\bibfnamefont{S.}~\bibnamefont{Ge}},
  \bibinfo{author}{\bibfnamefont{X.}~\bibnamefont{Li}},
  \bibinfo{author}{\bibfnamefont{J.}~\bibnamefont{Qiu}},
  \bibinfo{author}{\bibfnamefont{Y.}~\bibnamefont{Ji}},
  \bibinfo{author}{\bibfnamefont{J.}~\bibnamefont{Feng}}, \bibnamefont{and}
  \bibinfo{author}{\bibfnamefont{D.}~\bibnamefont{Sun}}, \bibinfo{journal}{ACS
  Nano} \textbf{\bibinfo{volume}{7}}, \bibinfo{pages}{11087}
  (\bibinfo{year}{2013}).

\bibitem[{\citenamefont{Mai et~al.}(2014)\citenamefont{Mai, Barrette, Yu,
  Semenov, Kim, Cao, and Gundogdu}}]{Mai:2014a}
\bibinfo{author}{\bibfnamefont{C.}~\bibnamefont{Mai}},
  \bibinfo{author}{\bibfnamefont{A.}~\bibnamefont{Barrette}},
  \bibinfo{author}{\bibfnamefont{Y.}~\bibnamefont{Yu}},
  \bibinfo{author}{\bibfnamefont{Y.~G.} \bibnamefont{Semenov}},
  \bibinfo{author}{\bibfnamefont{K.~W.} \bibnamefont{Kim}},
  \bibinfo{author}{\bibfnamefont{L.}~\bibnamefont{Cao}}, \bibnamefont{and}
  \bibinfo{author}{\bibfnamefont{K.}~\bibnamefont{Gundogdu}},
  \bibinfo{journal}{Nano Letters} \textbf{\bibinfo{volume}{14}},
  \bibinfo{pages}{202} (\bibinfo{year}{2014}).

\bibitem[{\citenamefont{Wang et~al.}(2014{\natexlab{b}})\citenamefont{Wang,
  Bouet, Lagarde, Vidal, Balocchi, Amand, Marie, and Urbaszek}}]{Wang:2014b}
\bibinfo{author}{\bibfnamefont{G.}~\bibnamefont{Wang}},
  \bibinfo{author}{\bibfnamefont{L.}~\bibnamefont{Bouet}},
  \bibinfo{author}{\bibfnamefont{D.}~\bibnamefont{Lagarde}},
  \bibinfo{author}{\bibfnamefont{M.}~\bibnamefont{Vidal}},
  \bibinfo{author}{\bibfnamefont{A.}~\bibnamefont{Balocchi}},
  \bibinfo{author}{\bibfnamefont{T.}~\bibnamefont{Amand}},
  \bibinfo{author}{\bibfnamefont{X.}~\bibnamefont{Marie}}, \bibnamefont{and}
  \bibinfo{author}{\bibfnamefont{B.}~\bibnamefont{Urbaszek}},
  \bibinfo{journal}{Phys. Rev. B} \textbf{\bibinfo{volume}{90}},
  \bibinfo{pages}{075413} (\bibinfo{year}{2014}{\natexlab{b}}).

\bibitem[{\citenamefont{{Rivera} et~al.}(2014)\citenamefont{{Rivera},
  {Schaibley}, {Jones}, {Ross}, {Wu}, {Aivazian}, {Klement}, {Ghimire}, {Yan},
  {Mandrus} et~al.}}]{Rivera:2014a}
\bibinfo{author}{\bibfnamefont{P.}~\bibnamefont{{Rivera}}},
  \bibinfo{author}{\bibfnamefont{J.~R.} \bibnamefont{{Schaibley}}},
  \bibinfo{author}{\bibfnamefont{A.~M.} \bibnamefont{{Jones}}},
  \bibinfo{author}{\bibfnamefont{J.~S.} \bibnamefont{{Ross}}},
  \bibinfo{author}{\bibfnamefont{S.}~\bibnamefont{{Wu}}},
  \bibinfo{author}{\bibfnamefont{G.}~\bibnamefont{{Aivazian}}},
  \bibinfo{author}{\bibfnamefont{P.}~\bibnamefont{{Klement}}},
  \bibinfo{author}{\bibfnamefont{N.~J.} \bibnamefont{{Ghimire}}},
  \bibinfo{author}{\bibfnamefont{J.}~\bibnamefont{{Yan}}},
  \bibinfo{author}{\bibfnamefont{D.~G.} \bibnamefont{{Mandrus}}},
  \bibnamefont{et~al.}, \bibinfo{journal}{ArXiv e-prints}
  (\bibinfo{year}{2014}), \eprint{1403.4985}.

\bibitem[{\citenamefont{{Yu} et~al.}(2014)\citenamefont{{Yu}, {Hu}, {Su},
  {Huang}, {Liu}, {Jin}, {Purezky}, {Geohegan}, {Kim}, {Zhang}
  et~al.}}]{Yu:2014d}
\bibinfo{author}{\bibfnamefont{Y.}~\bibnamefont{{Yu}}},
  \bibinfo{author}{\bibfnamefont{S.}~\bibnamefont{{Hu}}},
  \bibinfo{author}{\bibfnamefont{L.}~\bibnamefont{{Su}}},
  \bibinfo{author}{\bibfnamefont{L.}~\bibnamefont{{Huang}}},
  \bibinfo{author}{\bibfnamefont{Y.}~\bibnamefont{{Liu}}},
  \bibinfo{author}{\bibfnamefont{Z.}~\bibnamefont{{Jin}}},
  \bibinfo{author}{\bibfnamefont{A.~A.} \bibnamefont{{Purezky}}},
  \bibinfo{author}{\bibfnamefont{D.~B.} \bibnamefont{{Geohegan}}},
  \bibinfo{author}{\bibfnamefont{K.~W.} \bibnamefont{{Kim}}},
  \bibinfo{author}{\bibfnamefont{Y.}~\bibnamefont{{Zhang}}},
  \bibnamefont{et~al.}, \bibinfo{journal}{ArXiv e-prints}
  (\bibinfo{year}{2014}), \eprint{1403.6181}.

\bibitem[{\citenamefont{Urbaszek et~al.}(2013)\citenamefont{Urbaszek, Marie,
  Amand, Krebs, Voisin, Maletinsky, H\"ogele, and Imamoglu}}]{Urbaszek:2013a}
\bibinfo{author}{\bibfnamefont{B.}~\bibnamefont{Urbaszek}},
  \bibinfo{author}{\bibfnamefont{X.}~\bibnamefont{Marie}},
  \bibinfo{author}{\bibfnamefont{T.}~\bibnamefont{Amand}},
  \bibinfo{author}{\bibfnamefont{O.}~\bibnamefont{Krebs}},
  \bibinfo{author}{\bibfnamefont{P.}~\bibnamefont{Voisin}},
  \bibinfo{author}{\bibfnamefont{P.}~\bibnamefont{Maletinsky}},
  \bibinfo{author}{\bibfnamefont{A.}~\bibnamefont{H\"ogele}}, \bibnamefont{and}
  \bibinfo{author}{\bibfnamefont{A.}~\bibnamefont{Imamoglu}},
  \bibinfo{journal}{Rev. Mod. Phys.} \textbf{\bibinfo{volume}{85}},
  \bibinfo{pages}{79} (\bibinfo{year}{2013}).

\bibitem[{\citenamefont{Sahin et~al.}(2013)\citenamefont{Sahin, Tongay, Horzum,
  Fan, Zhou, Li, Wu, and Peeters}}]{Sahin:2013a}
\bibinfo{author}{\bibfnamefont{H.}~\bibnamefont{Sahin}},
  \bibinfo{author}{\bibfnamefont{S.}~\bibnamefont{Tongay}},
  \bibinfo{author}{\bibfnamefont{S.}~\bibnamefont{Horzum}},
  \bibinfo{author}{\bibfnamefont{W.}~\bibnamefont{Fan}},
  \bibinfo{author}{\bibfnamefont{J.}~\bibnamefont{Zhou}},
  \bibinfo{author}{\bibfnamefont{J.}~\bibnamefont{Li}},
  \bibinfo{author}{\bibfnamefont{J.}~\bibnamefont{Wu}}, \bibnamefont{and}
  \bibinfo{author}{\bibfnamefont{F.~M.} \bibnamefont{Peeters}},
  \bibinfo{journal}{Phys. Rev. B} \textbf{\bibinfo{volume}{87}},
  \bibinfo{pages}{165409} (\bibinfo{year}{2013}).

\bibitem[{\citenamefont{Zhao et~al.}(2013{\natexlab{a}})\citenamefont{Zhao,
  Ghorannevis, Chu, Toh, Kloc, Tan, and Eda}}]{Zhao:2013b}
\bibinfo{author}{\bibfnamefont{W.}~\bibnamefont{Zhao}},
  \bibinfo{author}{\bibfnamefont{Z.}~\bibnamefont{Ghorannevis}},
  \bibinfo{author}{\bibfnamefont{L.}~\bibnamefont{Chu}},
  \bibinfo{author}{\bibfnamefont{M.}~\bibnamefont{Toh}},
  \bibinfo{author}{\bibfnamefont{C.}~\bibnamefont{Kloc}},
  \bibinfo{author}{\bibfnamefont{P.-H.} \bibnamefont{Tan}}, \bibnamefont{and}
  \bibinfo{author}{\bibfnamefont{G.}~\bibnamefont{Eda}}, \bibinfo{journal}{ACS
  Nano} \textbf{\bibinfo{volume}{7}}, \bibinfo{pages}{791}
  (\bibinfo{year}{2013}{\natexlab{a}}).

\bibitem[{\citenamefont{Zhao et~al.}(2013{\natexlab{b}})\citenamefont{Zhao,
  Ribeiro, Toh, Carvalho, Kloc, Castro~Neto, and Eda}}]{Zhao:2013c}
\bibinfo{author}{\bibfnamefont{W.}~\bibnamefont{Zhao}},
  \bibinfo{author}{\bibfnamefont{R.~M.} \bibnamefont{Ribeiro}},
  \bibinfo{author}{\bibfnamefont{M.}~\bibnamefont{Toh}},
  \bibinfo{author}{\bibfnamefont{A.}~\bibnamefont{Carvalho}},
  \bibinfo{author}{\bibfnamefont{C.}~\bibnamefont{Kloc}},
  \bibinfo{author}{\bibfnamefont{A.~H.} \bibnamefont{Castro~Neto}},
  \bibnamefont{and} \bibinfo{author}{\bibfnamefont{G.}~\bibnamefont{Eda}},
  \bibinfo{journal}{Nano Letters} \textbf{\bibinfo{volume}{13}},
  \bibinfo{pages}{5627} (\bibinfo{year}{2013}{\natexlab{b}}).

\bibitem[{\citenamefont{Debbichi et~al.}(2014)\citenamefont{Debbichi, Eriksson,
  and Leb\`egue}}]{Debbichi:2014a}
\bibinfo{author}{\bibfnamefont{L.}~\bibnamefont{Debbichi}},
  \bibinfo{author}{\bibfnamefont{O.}~\bibnamefont{Eriksson}}, \bibnamefont{and}
  \bibinfo{author}{\bibfnamefont{S.}~\bibnamefont{Leb\`egue}},
  \bibinfo{journal}{Phys. Rev. B} \textbf{\bibinfo{volume}{89}},
  \bibinfo{pages}{205311} (\bibinfo{year}{2014}).

\bibitem[{typ()}]{type2}
\bibinfo{note}{The calculations of \cite{Yu:2014c} suggest that the low energy
  PL transitions is indirect in real space (type II), not k-space.}

\bibitem[{\citenamefont{Harle et~al.}(1994)\citenamefont{Harle, Bolay, Lux,
  Scholz, Michler, Moritz, Forner, and Hangleiter}}]{Harley:1994a}
\bibinfo{author}{\bibfnamefont{V.}~\bibnamefont{Harle}},
  \bibinfo{author}{\bibfnamefont{H.}~\bibnamefont{Bolay}},
  \bibinfo{author}{\bibfnamefont{E.}~\bibnamefont{Lux}},
  \bibinfo{author}{\bibfnamefont{F.}~\bibnamefont{Scholz}},
  \bibinfo{author}{\bibfnamefont{P.}~\bibnamefont{Michler}},
  \bibinfo{author}{\bibfnamefont{A.}~\bibnamefont{Moritz}},
  \bibinfo{author}{\bibfnamefont{T.}~\bibnamefont{Forner}}, \bibnamefont{and}
  \bibinfo{author}{\bibfnamefont{A.}~\bibnamefont{Hangleiter}}, in
  \emph{\bibinfo{booktitle}{Indium Phosphide and Related Materials, 1994.
  Conference Proceedings., Sixth International Conference on}}
  (\bibinfo{year}{1994}), pp. \bibinfo{pages}{6--9}.

\bibitem[{\citenamefont{{Riley} et~al.}(2014)\citenamefont{{Riley}, {Mazzola},
  {Dendzik}, {Michiardi}, {Takayama}, {Bawden}, {Graner{\o}d}, {Leandersson},
  {Balasubramanian}, {Hoesch} et~al.}}]{Riley:2014a}
\bibinfo{author}{\bibfnamefont{J.~M.} \bibnamefont{{Riley}}},
  \bibinfo{author}{\bibfnamefont{F.}~\bibnamefont{{Mazzola}}},
  \bibinfo{author}{\bibfnamefont{M.}~\bibnamefont{{Dendzik}}},
  \bibinfo{author}{\bibfnamefont{M.}~\bibnamefont{{Michiardi}}},
  \bibinfo{author}{\bibfnamefont{T.}~\bibnamefont{{Takayama}}},
  \bibinfo{author}{\bibfnamefont{L.}~\bibnamefont{{Bawden}}},
  \bibinfo{author}{\bibfnamefont{C.}~\bibnamefont{{Graner{\o}d}}},
  \bibinfo{author}{\bibfnamefont{M.}~\bibnamefont{{Leandersson}}},
  \bibinfo{author}{\bibfnamefont{T.}~\bibnamefont{{Balasubramanian}}},
  \bibinfo{author}{\bibfnamefont{M.}~\bibnamefont{{Hoesch}}},
  \bibnamefont{et~al.}, \bibinfo{journal}{ArXiv e-prints}
  (\bibinfo{year}{2014}), \eprint{1408.6778}.

\bibitem[{\citenamefont{Zhang et~al.}(2014{\natexlab{b}})\citenamefont{Zhang,
  Liu, Luo, Freeman, and Zunger}}]{Zhang:2014b}
\bibinfo{author}{\bibfnamefont{X.}~\bibnamefont{Zhang}},
  \bibinfo{author}{\bibfnamefont{Q.}~\bibnamefont{Liu}},
  \bibinfo{author}{\bibfnamefont{J.-W.} \bibnamefont{Luo}},
  \bibinfo{author}{\bibfnamefont{A.~J.} \bibnamefont{Freeman}},
  \bibnamefont{and} \bibinfo{author}{\bibfnamefont{A.}~\bibnamefont{Zunger}},
  \bibinfo{journal}{Nat. Physics} \textbf{\bibinfo{volume}{10}},
  \bibinfo{pages}{387} (\bibinfo{year}{2014}{\natexlab{b}}).

\bibitem[{imp()}]{imppol}
\bibinfo{note}{We record a small but not zero linear polarization for a
  detection energy $\approx 1.63$~eV, compare also with the shoulder in
  Fig.~\ref{fig:fig3}a on the $P_c$. Although not resolved in our PL spectrum,
  this energy was identified by Jones et al as impurity-bound excitons in
  WSe$_2$ bilayers \cite{Jones:2014a}.}

\bibitem[{\citenamefont{Glazov et~al.}(2014)\citenamefont{Glazov, Amand, Marie,
  Lagarde, Bouet, and Urbaszek}}]{Glazov:2014a}
\bibinfo{author}{\bibfnamefont{M.~M.} \bibnamefont{Glazov}},
  \bibinfo{author}{\bibfnamefont{T.}~\bibnamefont{Amand}},
  \bibinfo{author}{\bibfnamefont{X.}~\bibnamefont{Marie}},
  \bibinfo{author}{\bibfnamefont{D.}~\bibnamefont{Lagarde}},
  \bibinfo{author}{\bibfnamefont{L.}~\bibnamefont{Bouet}}, \bibnamefont{and}
  \bibinfo{author}{\bibfnamefont{B.}~\bibnamefont{Urbaszek}},
  \bibinfo{journal}{Phys. Rev. B} \textbf{\bibinfo{volume}{89}},
  \bibinfo{pages}{201302} (\bibinfo{year}{2014}).

\bibitem[{\citenamefont{Yu and Wu}(2014{\natexlab{b}})}]{Yu:2014a}
\bibinfo{author}{\bibfnamefont{T.}~\bibnamefont{Yu}} \bibnamefont{and}
  \bibinfo{author}{\bibfnamefont{M.~W.} \bibnamefont{Wu}},
  \bibinfo{journal}{Phys. Rev. B} \textbf{\bibinfo{volume}{89}},
  \bibinfo{pages}{205303} (\bibinfo{year}{2014}{\natexlab{b}}).

\bibitem[{\citenamefont{{Zhu} et~al.}(2014)\citenamefont{{Zhu}, {Zhang},
  {Glazov}, {Urbaszek}, {Amand}, {Ji}, {Liu}, and {Marie}}}]{Zhucr:2014a}
\bibinfo{author}{\bibfnamefont{C.~R.} \bibnamefont{{Zhu}}},
  \bibinfo{author}{\bibfnamefont{K.}~\bibnamefont{{Zhang}}},
  \bibinfo{author}{\bibfnamefont{M.}~\bibnamefont{{Glazov}}},
  \bibinfo{author}{\bibfnamefont{B.}~\bibnamefont{{Urbaszek}}},
  \bibinfo{author}{\bibfnamefont{T.}~\bibnamefont{{Amand}}},
  \bibinfo{author}{\bibfnamefont{Z.~W.} \bibnamefont{{Ji}}},
  \bibinfo{author}{\bibfnamefont{B.~L.} \bibnamefont{{Liu}}}, \bibnamefont{and}
  \bibinfo{author}{\bibfnamefont{X.}~\bibnamefont{{Marie}}},
  \bibinfo{journal}{ArXiv e-prints}  (\bibinfo{year}{2014}),
  \eprint{1407.5862}.

\bibitem[{\citenamefont{Fogler et~al.}(2014)\citenamefont{Fogler, Butov, and
  Novoselov}}]{Fogler:2014a}
\bibinfo{author}{\bibfnamefont{M.~M.} \bibnamefont{Fogler}},
  \bibinfo{author}{\bibfnamefont{L.~V.} \bibnamefont{Butov}}, \bibnamefont{and}
  \bibinfo{author}{\bibfnamefont{K.~S.} \bibnamefont{Novoselov}},
  \bibinfo{journal}{Nat Commun} \textbf{\bibinfo{volume}{5}},
  \bibinfo{pages}{10.1038/ncomms5555} (\bibinfo{year}{2014}).

\end{thebibliography}
\end{document}